

\input harvmac
\noblackbox
%
%

%
%

\def\Title#1#2{\ifx\answ\bigans \nopagenumbers
\abstractfont\hsize=\hstitle\rightline{#1}%
\vskip .5in\centerline{\titlefont #2}\abstractfont\vskip .5in\pageno=0
\else \rightline{#1}
\vskip .8in\centerline{\titlefont #2}
\vskip .5in\pageno=1\fi}
\ifx\answ\bigans

scaled\magstep3
\else

scaled\magstep3
 
 \font\absi=cmmi10 scaled\magstep1
\font\absis=cmmi7 scaled\magstep1 \font\absiss=cmmi5 scaled\magstep1
\font\abssy=cmsy10 scaled\magstep1 \font\abssys=cmsy7 scaled\magstep1
\font\abssyss=cmsy5 scaled\magstep1 
\skewchar\absi='177 \skewchar\absis='177 \skewchar\absiss='177
\skewchar\abssy='60 \skewchar\abssys='60 \skewchar\abssyss='60
\fi
%
%

\def\ajou#1&#2(#3){\ \sl#1\bf#2\rm(19#3)}

\def\frac#1#2{{#1 \over #2}}

\def\mn{{\mu\nu}}

\def\R{\hbox{\rm I \kern-5pt R}}
\def\ajou#1&#2(#3){\ \sl#1\bf#2\rm(19#3)}
\def\bA{\bar{A}}
\def\by{\bar{y}}
\def\bt{\bar{t}}

\def\l3{\lambda_3}
\def\bphi{\bar{\varphi}}
\def\brho{\bar{\rho}}

\hyphenation{par-am-et-rised}

%
%
\lref\dray{A. Ashtekar and T. Dray, \ajou Comm. Math. Phys. &79 (81) 581;
T. Dray, \ajou Gen. Rel. Grav. &14 (82) 109.}
\lref\dgt{H.F. Dowker, R. Gregory and J. Traschen, \ajou Phys. Rev. &D45 (92)
2762.}
\lref\sg{S. B. Giddings, ``Constraints on Black Hole Remnants",
Santa Barbara preprint UCSBTH-93-08, hep-th@xxx/9304027.}
\lref\wilc{C.F.E. Holzhey and F. Wilczek, \ajou Nucl.Phys. &B380 (92) 447.}
\lref\dggh{H.F. Dowker, J.P. Gauntlett, S. B. Giddings and G. Horowitz,
In Preparation.}
\lref\andy{D. Garfinkle, S. Giddings and A. Strominger, ``Entropy in Black
Hole Pair Production'', Santa Barbara preprint UCSBTH-93-17,
gr-qc/9306023.}
\lref\kw{W. Kinnersley and M. Walker, \ajou Phys. Rev. &D2 (70) 1359.}
\lref\senny{S. Hassan and A. Sen, \ajou Nucl. Phys. &B375 (92) 103.}
\lref\he{J. Ehlers, in {\it Les Theories de la Gravitation},
(CNRS, Paris, 1959); B. Harrison, \ajou J. Math. Phys. &9 (68) 1744.}
\lref\cghs{C.G. Callan, S.B. Giddings, J.A. Harvey and A. Strominger,
\ajou Phys. Rev. &D45 (92) R1005.}
\lref\ebh{S.B. Giddings and A. Strominger, \ajou Phys. Rev. &D46 (92) 627.}
\lref\kkbh{H. Leutwyler, \ajou Arch. Sci. &13 (60) 549;
P. Dobiasch and D. Maison, \ajou Gen. Rel. and Grav. &14 (82) 231;
A. Chodos and S. Detweiler, \ajou Gen. Rel. and Grav. &14 (82) 879;
D. Pollard, \ajou J. Phys. A &16 (83) 565;
G.W. Gibbons and D.L. Wiltshire, \ajou Ann.Phys. &167
(86) 201;
erratum \ajou ibid. &176 (87) 393.}
\lref\gm{G.W. Gibbons and K. Maeda,
\ajou Nucl. Phys. &B298
(88) 741.}
\lref\ghs{D. Garfinkle, G. Horowitz, and A. Strominger,
\ajou Phys. Rev. &D43 (91) 3140, erratum\ajou Phys. Rev.
& D45 (92) 3888.}
\lref\AfMa{I.K. Affleck and N.S. Manton,
\ajou Nucl. Phys. &B194 (82) 38.}
\lref\AAM{I.K. Affleck, O. Alvarez, and N.S. Manton,
\ajou Nucl. Phys. &B197 (82) 509.}
\lref\geroch{R.P. Geroch, \ajou J. Math. Phys. &8 (67) 782.}
\lref\raftop{R.D. Sorkin, in
{\sl Proceedings of the Third Canadian Conference on General
Relativity and Relativistic Astrophysics}, (Victoria, Can\-ada, May 1989),
eds. A. Coley, F. Cooperstock and B. Tupper (World Scientific, 1990).}
\lref\wheeler{J. Wheeler, \ajou Ann. Phys. &2 (57) 604.}
\lref\gwg{G.W. Gibbons,
in {\sl Fields and geometry}, proceedings of
22nd Karpacz Winter School of Theoretical Physics: Fields and
Geometry, Karpacz, Poland, Feb 17 - Mar 1, 1986, ed. A. Jadczyk (World
Scientific, 1986).}
\lref\garstrom{D. Garfinkle and A. Strominger,
\ajou Phys. Lett. &256B (91) 146.}
\lref\ernst{F. J. Ernst, \ajou J. Math. Phys. &17 (76) 515.}
\lref\rafKK{R. Sorkin, \ajou Phys. Rev. Lett. &51 (83) 87.}
\lref\grossperry{D. Gross and M.J. Perry, \ajou Nucl.Phys. &B226 (83) 29.}
\lref\hawk{S.W. Hawking in {\sl General relativity : an Einstein centenary
survey}, eds. S.W. Hawking, W. Israel (Cambridge University Press, Cambridge,
New York, 1979).}
\lref\schwinger{J. Schwinger}
\lref\melvin{M. A. Melvin, \ajou Phys. Lett. &8 (64) 65.}
\lref\banks{T. Banks, A. Dhabolkar, M.R. Douglas and M. O'Loughlin
\ajou Phys.Rev. &D45 (92) 3607.}
\lref\banksol{T. Banks, M. O'Loughlin and A. Strominger, \ajou Phys.Rev.
&D47 (93) 4476.}
\lref\ginsperry{P. Ginsparg and M.J. Perry, \ajou Nucl.Phys. &B222 (83) 245.}
\lref\MATHEMATICA{
Wolfram Research, Inc., {\rm MATHEMATICA} (Wolfram Research, Inc.,
Champaign, Illinois, 1992).}
\lref\MATHTENSOR{
L. Parker and S. M Christensen, {\rm MATHTENSOR} (MathSolutions, Inc., Chapel
Hill, North Carolina, 1992).}

%
%
\Title{\vbox{\baselineskip12pt
\hbox{EFI-93-51}
\hbox{FERMILAB-Pub-93/272-A}
\hbox{UMHEP-393}
\hbox{hep-th/9309075}}}
{\vbox{\centerline{Pair Creation of Dilaton Black Holes }
       }}
{
\baselineskip=12pt
\centerline{Fay Dowker,$^1$\footnote*{Address from 1st Oct. 1993: Relativity
 Group
Department of Physics, University of California, Santa Barbara, CA 93106.}
 Jerome P. Gauntlett,$^2$ David A. Kastor,$^{3a}$
Jennie Traschen$^{3b}$}
\bigskip
\centerline{\sl $^1$NASA/Fermilab Astrophysics Group}
\centerline{\sl Fermi National Accelerator Laboratory}
\centerline{\sl PO Box 500, Batavia, IL 60510}
\bigskip
\centerline{\sl $^2$Enrico Fermi Institute, University of Chicago}
\centerline{\sl 5640 Ellis Avenue, Chicago, IL 60637 }
\centerline{\it Internet: jerome@yukawa.uchicago.edu}
\bigskip
\centerline{\sl $^3$Department of Physics and Astronomy}
\centerline{\sl University of Massachusetts}
\centerline{\sl Amherst, MA 01003-4525}
\centerline{\it $^a$Internet: kastor@phast.umass.edu}
\centerline{\it $^b$Internet: lboo@phast.umass.edu}
\medskip
\centerline{\bf Abstract}
We consider dilaton gravity theories in four spacetime dimensions
parametrised by a constant $a$, which controls the dilaton coupling,
and construct new exact solutions. We first
generalise the C-metric of Einstein-Maxwell theory ($a=0$) to
solutions corresponding to oppositely charged dilaton black holes
undergoing uniform acceleration for general $a$.
We next develop a solution generating technique which allows us
to ``embed" the dilaton C-metrics in magnetic dilaton Melvin
backgrounds, thus generalising the Ernst metric of Einstein-Maxwell theory.
By adjusting the parameters appropriately, it is possible to
eliminate the nodal singularities of the dilaton C-metrics.
For $a<1$ (but not for $a\ge 1$), it is possible to further restrict
the parameters so that the dilaton Ernst solutions have a smooth euclidean
section with topology $S^2\times S^2-{\rm\{pt\}}$, corresponding
to instantons describing the pair production of dilaton black holes
in a magnetic field. A different restriction on the parameters leads to
smooth instantons for all values of $a$ with topology $S^2\times \R^2$.}



%
%

\newsec{Introduction}
The idea that the topology of space might change in a quantum
theory of gravity is
an old one \wheeler.
The ``canonical'' approach to quantum gravity, however, rules out the
possibility from the start by  taking the configuration space to be
the space of three-geometries on a {\it fixed} three-manifold and
the ``covariant'' approach assumes a fixed background {\it spacetime}.
Thus, the most natural framework for quantum gravity in which
to investigate
topology changing processes seems to be the sum-over-histories.
In the sum-over-histories
formulation a  topology changing transition amplitude is given by a
functional integral over four-metrics on four-manifolds (cobordisms)
with boundaries which agree with the initial and final states. What
conditions to place on the metrics summed over is a matter for some
debate. One approach is to only sum over
euclidean metrics \hawk. Another proposal is to sum over  almost
everywhere lorentzian metrics, restricting the metrics to be
causality preserving (i.e. no
closed time-like curves),
in which case the issue of the necessary singularities must be
broached \raftop.  Although such functional integrals
are ill-defined as yet, one can still do calculations by
assuming that they
can be well approximated by saddle point methods.
An instanton, a euclidean solution that
interpolates between the initial and final states
of a classically forbidden transition,
is a saddle point for both the ``euclidean'' and ``lorentzian''
functional integrals.
We take the existence of an instanton
as an indication that the transition has a  finite rate and must be
taken into consideration.

One such instanton in Einstein-Maxwell
theory is the euclideanised Ernst metric
\ernst\ which is interpreted as describing the pair production of
two magnetically charged Reissner-Nordstrom
black holes in a Melvin magnetic universe \refs{\gwg, \garstrom}. This is the
gravitational analogue of the Schwinger pair production of charged
particles in a uniform electromagnetic field.
In this process the topology of space changes from $\R^3$ to
$S^2 \times S^1\, -  \{{\rm pt}\}$
corresponding to the formation of two oppositely charged black holes
whose throats are connected by a handle.
The calculation of the rate of this process leads to the observation
that it is enhanced over the production rate of monopoles by a
factor $e^{S_{bh}}$ where $S_{bh}$ is the Hawking-Bekenstein
entropy of the black holes \refs{\andy}. This supports the notion that the
entropy counts the number of ``internal'' states of the black hole.

Are similar processes described by instantons in other theories
containing gravity?
It is known, for example, that both the
low energy limit of string theory \refs{\gm, \ghs} and 5-dimensional
Kaluza-Klein theory \refs{\rafKK, \grossperry, \kkbh}
admit a family of charged black hole solutions.
One may ask if instantons exist which describe their
pair production.
An action which includes all the above mentioned theories describes
the interaction between a dilaton, a $U(1)$ gauge
field and gravity and is given by
\eqn\action{
S=\int d^4x {\sqrt {-g}}\left[R-2(\nabla\phi)^2-e^{-2a\phi}F^2\right].
}
For $a=0$ this is just standard Einstein-Maxwell theory. For $a=1$
it is a part of the action describing the low-energy dynamics of
string theory, while for $a=\sqrt 3$ it
arises from 5-dimensional Kaluza-Klein theory. For each value of $a$,
there exists a two parameter family of black hole solutions
(which we shall briefly review in section 2)
labelled by the mass $m$ and the magnetic (or electric) charge $q$.
That topology changing instantons for \action\ exist, at least for
$a<1$, describing
the pair creation of such black holes, will be one
of the results of this paper.

It is important to note that we have used the ``Einstein" metric
to describe these theories. Metrics rescaled by a dilaton dependent
factor are also of physical interest and may have different causal
structures.
For example, in string theory ($a=1$) the ``sigma-model"
metric $g_\sigma=e^{2\phi}g$ is the metric that couples to the
string degrees of freedom.
If we consider the magnetically charged black holes in this theory, then for
$m>{\sqrt 2}q$ both the
Einstein metric and the sigma model metric
have a singularity cloaked by an event horizon. However, in the
extremal limit, $m={\sqrt 2}q$, the Einstein metric
has a naked singularity, whereas in the sigma
model metric the singularity disappears from the
space-time, down an infinitely long tube.
In this limit
the sigma model metric is geodesically complete and moreover the upper
bound on the curvature can be made as small as one likes by choosing
$q$ large enough.

These properties of the sigma model metric
are part of the motivation for using the $a=1$
theory to further understand the issue of information loss
in the scattering of matter with extremal black holes.
The low-energy scattering of particles with such an extremal black hole,
including the effects of
back reaction on the metric, has been studied in \refs{\cghs,\ebh,\banks}.
One truncates to the s-wave sector of the theory and considers an effective
two-dimensional theory defined in the throat region.
Using
semi-classical techniques, it has been argued that there may exist an
infinite number of near degenerate states corresponding to massless
modes propagating down the throat.
It was conjectured in \refs{\cghs,\ebh,\banks}
that these remnants or ``cornucopions" are the end-points of
Hawking evaporation. The infinite length of the throat allows
for an arbitrarily large number of remnants, which can then
store an arbitrarily large amount of information.

One objection to this scenario is that if an infinite number
of such remnants exist, then we may expect them to each have a
finite probability of being pair created. The infinite number
of species would then lead to divergences in ordinary quantum field theory
processes. A way around this objection was proposed in
\banksol , where the rate of production of these remnants in a magnetic field
was estimated using instanton methods. It was argued that the rate
of pair production is not infinite, because the instanton would
produce a pair of throats connected by a finite length handle.
The finite length of the throat would then imply that only
a finite number of remnants could be excited and that the total
production rate
would be finite. A shortcoming of the arguments in \banksol , however,
was that no exact instanton solutions were constructed. Looking for such
exact solutions was one of the motivations for the present work.

The plan of the rest of the paper is as follows.
In section 2 we present the dilaton generalisations of the C-metric for
arbitrary dilaton coupling $a$.
These describe two oppositely charged
dilaton black holes accelerating away from each other.
We show that, just as in the Einstein-Maxwell C-metric, there exist
nodal singularities in the metric which cannot be removed by any choice
of period for the azimuthal coordinate. These can be thought of as providing
the forces necessary to accelerate the black holes.
In Einstein-Maxwell theory,
string theory and Kaluza-Klein theory there
are known
transformations which generate new solutions starting
from a known static, axisymmetric
solution. In section 3, we show that such generating transformations
exist for all $a$, and take flat space into
dilaton magnetic Melvin universes.
When applied to the C-metrics, these same
transformations give dilaton generalisations of the Ernst solution;
choosing the parameters appropriately, the magnetic field
can provide exactly the right amount of acceleration to remove the nodal
singularities.
In section 4, we discuss the euclidean section of the dilaton Ernst
solutions. To obtain a regular geometry, it is necessary
that the Hawking temperatures of the black hole and acceleration
horizons be equal. For $a<1$, we find that
it is possible to do this at non-zero
temperature, and one obtains natural generalisations of the $q=m$ instantons
discussed in \refs{\gwg, \garstrom} with
topology $S^2\times S^2 -{\rm\{pt\}}$.
These instantons describe the formation of a Wheeler wormhole on a
spatial slice of a magnetic dilaton Melvin universe.  For all values of $a$,
it is
possible to obtain a smooth euclidean section in the limit that
the two horizons have zero temperature. These instantons
have topology $S^2\times \R^2$. The physical interpretation of these
instantons, however, is unclear.
Section 5 is a summary and discussion of our
results.

\newsec{Dilaton C-metrics}
\subsec{Charged Black Holes in Dilaton Gravity}

The equations of motion coming from the action \action\ are given by
\eqn\eom{
\eqalign{
&\nabla_\mu(e^{-2a\phi}F^{\mn})=0\cr
&\nabla^2\phi+{a\over 2}e^{-2a\phi}F^2=0\cr
&R_\mn=2\nabla_\mu\phi\nabla_\nu\phi+2e^{-2a\phi}F_{\mu\rho}F^{\,\rho}_\nu
-{1\over 2}g_\mn e^{-2a\phi}F^2.\cr}
}
These equations are invariant with respect to an electric-magnetic duality
transformation, under which the metric is
unchanged and the new field strength
$\tilde{F}$ and dilaton $\tilde{\phi}$ are given by
\eqn\dual{
\tilde{F}_{\mu\nu} = \half e^{-2a\phi}\epsilon_{\mu\nu\rho\sigma}
F^{\rho\sigma},  \qquad \tilde{\phi}=-\phi .
}
We will only consider the magnetically charged solutions below, but because
of this duality our results also apply to the electric case.

For given $a$ the equations of motion \eom\ admit a two parameter family
of magnetically charged black hole solutions given by
\foot{To obtain solutions where the dilaton asymptotically approaches an
arbitrary constant $\phi_0$, one can
use the fact that the action is invariant
under $\phi\to\phi+\phi_0$, $F\to e^{a\phi_0}F$
and the metric left unchanged.
We will suppress $\phi_0$ in the following.} \gm\ghs\
\eqn\dbhs{
\eqalign{
& ds^2=-\lambda^2dt^2+\lambda^{-2}dr^2+R^2(d\theta^2 +
\sin^2\theta d\varphi^2)
 \cr
& e^{-2a\phi}=\left(1-{r_-\over r}\right)^{2a^2\over(1+a^2)},\qquad
\qquad A_\varphi=q{\rm cos}\theta\cr
&\lambda^2=\left(1-{r_+\over r}\right)
\left(1-{r_-\over r}\right)^{{(1-a^2)\over (1+a^2)}},\qquad
R^2=r^2\left(1-{r_-\over r}\right)^{2a^2\over (1+a^2)}.\cr}
}
Assuming $r_+ > r_-$, then $r_+$ is the location of a black hole horizon.
For $a=0$, $r_-$ is the location of the
inner Cauchy horizon, however for $a>0$
the surface $r=r_-$ is singular.
The parameters $r_+$ and $r_-$ are related to the ADM mass $m$
and total charge $q$ by
\eqn\mass{
m={r_+\over 2} + \left ({1-a^2\over 1+a^2}\right ){r_-\over 2},\qquad
q=\left({r_+r_-\over 1+a^2}\right)^{1\over 2}.
}
The extremal limit occurs when $r_+=r_-$.

Following \gm\ we introduce the ``total metric", $ds_T^2$, defined
via a conformal rescaling of the ``Einstein" metric
\eqn\total{ds_T^2=e^{2\phi/a}ds^2.}
For certain values of $a$ this metric naturally appears
in Kaluza-Klein theories \gm.
For $a=1$ this is just the sigma model metric that couples to the
string degrees of freedom. For $a<1$, in the extremal limit,
the total metric is geodesically complete and the
spatial sections have the form of two asymptotic regions joined by a
wormhole, one region being flat, the other having a deficit solid
angle.
For $a=1$ the geometry is that of an infinitely long throat.

\subsec{Dilaton C-Metric}

In Einstein-Maxwell theory the C-metric can be interpreted as the spacetime
corresponding to two Reissner-Nordstrom black holes
of opposite charge undergoing uniform acceleration \kw. The generalisation
of this spacetime to dilaton gravity is given by
\eqn\C{
\eqalign{
&ds^2={1\over A^2(x-y)^2}\left[F(x)\left\{G(y)dt^2-G^{-1}(y)dy^2\right\}
+F(y)\left\{G^{-1}(x)dx^2+G(x)d\varphi^2\right\}\right]\cr
&e^{-2a\phi}={F(y)\over F(x)},\qquad
A_\varphi=qx, \qquad F(\xi)=(1+r_-A\xi)^{2a^2\over (1+a^2)}\cr
&G(\xi)=\bar{G}(\xi)(1+r_-A\xi)^{(1-a^2)\over (1+a^2)},\qquad
\bar{G}(\xi)=\left[1-\xi^2(1+r_+A\xi)\right].\cr}
}
Note that  the form of $G$ as a product of two terms is quite similar to the
form of $\lambda$ in \dbhs\ and further, $\bar{G}$ is the cubic which appears
in the uncharged C-metric in \kw. The parameters
$q$, $r_-$ and $r_+$ are related as in
\mass.

The metric \C\ can be shown to give various known metrics in the
appropriate limits.  Setting $r_-=0$ gives the uncharged C-metric
(a vacuum solution and independent of $a$). Setting $a=0$ gives the charged
 C-metric
of Einstein-Maxwell theory, but in a slightly non-standard form:
the function $G$ is a quartic
with a linear term. To compare with the form of the C-metric given
in \kw \ one needs to
change coordinates to obtain a quartic with
no linear term. The appropriate transformations are discussed in \kw .

In the limit of zero acceleration, the metric \C\ reduces to the metric \dbhs\
for a single charged dilaton black hole.
To see this, it is useful to use new coordinates given by
\eqn\coord{
r=-{1\over Ay},\qquad T =A^{-1}t.
}
In these coordinates the metric \C\ becomes
\eqn\Ctwo{
\eqalign{
ds^2&={1\over (1+Arx)^2}\big[F(x)\left\{-H(r)dT^2 + H^{-1}(r)dr^2\right\}\cr
&{}\qquad\qquad\qquad+R^2(r)
\left\{G^{-1}(x)dx^2+G(x)d\varphi^2\right\}\big]\cr
H(r)&=(1-{r_+\over r}-A^2r^2)(1-{r_-\over r})^{1-a^2\over (1+a^2)},\cr}
}
where the function $R(r)$ is the same as that appearing in \dbhs .
Setting $A=0$ and $x={\rm cos}\theta$, we return to the metric \dbhs\
of the dilaton black holes.

The metric \C\ has two Killing vectors,
${\partial\over\partial t}$ and ${\partial\over\partial\varphi }$.
For the range of parameters $r_+A<2/(3\sqrt{3})$, the function $\bar{G}(\xi)$
has three real roots.  Denote these in ascending order by
$\xi_2$, $\xi_3$ and $\xi_4$ and define $\xi_1\equiv -{1\over r_-A}$.
One can show $\xi_3<\xi_4$ and we further restrict the parameters
so that $\xi_1 < \xi_2$.
The surface $y=\xi_1$ is
singular for $a>0$, as can be seen from the square of the field strength.
This surface is analogous to the singular surface
(the `would be' inner horizon) of the dilaton black holes.
The surface $y=\xi_2$ is the black hole horizon and the surface
$y=\xi_3$ is the acceleration horizon; they are both Killing horizons for
${\partial\over\partial t}$.

The coordinates $(x,\varphi)$ in \C\ are angular coordinates.
To keep the signature of the metric fixed, the coordinate $x$
is restricted to the range $\xi_3 \le x\le\xi_4$ in which $G(x)$ is positive.
The norm of the
Killing vector ${\partial\over\partial\varphi }$ vanishes at $x=\xi_3$ and
$x=\xi_4$, which correspond to the poles of spheres surrounding the black
holes.   The axis $x=\xi_3$ points along the symmetry axis
towards spatial infinity.  The axis $x=\xi_4$ points towards the other black
hole\foot{The coordinates we are using only cover the region of
spacetime where one of the black holes is.}.
Spatial infinity is reached by fixing $t$ and
letting both $y$ and $x$ approach
$\xi_3$.  Letting $y\rightarrow x$ for $x\ne\xi_3$ gives null or
timelike infinity \dray .
Since $y\rightarrow x$ is infinity, the range of the coordinate $y$ is
$-\infty<y<x$ for $a=0$, $\xi_1<y<x$ for $a>0$.

\subsec{Nodal Singularities}

As is the case with the ordinary C-metric,
it is not generally possible to choose the range of $\varphi$ such that the
metric \C\ is regular at both $x=\xi_3$ and $x=\xi_4$.
In order to see this, in a neighbourhood of each root
define a new coordinate $\theta$ according to
\eqn\newcoord{
\theta =\int^x_{\xi_i}{dx^{\prime}\over\sqrt{G(x^\prime)}} .
}
The angular part of the metric near one the poles $x=\xi_i$, $i=3,4$ then has
the form
\eqn\angular{
dl^2\approx {F(y)\over A^2(\xi_i-y)^2} \left(
d\theta^2 + {1\over 4}\lambda_i^2
\theta^2d\varphi ^2 \right ),
}
where $\lambda_i= |G^\prime(\xi_i)|$ and
one can show that
$\lambda_3<\lambda_4$.
Let the range of $\varphi$ be $0<\varphi\le\alpha$, then
the deficit angles at the two poles $\delta_3$, $\delta_4$ are given by
\eqn\node{\delta_3=2\pi - \half\alpha\lambda_3
\qquad \delta_4=2\pi - \half\alpha\lambda_4
}
We can remove the nodal singularity at $x=\xi_4$ by choosing
$\alpha=4\pi/\lambda_4$, but then there
is a positive deficit angle running along
the $\xi_3$ direction.  This corresponds to the black holes being pulled
by ``cosmic strings" of positive mass per unit length
$\mu=1-\lambda_3/\lambda_4$.
Alternatively, we can choose $\alpha=4\pi/\lambda_3$
to remove the nodal singularity at $\xi_3$.  This means there is a
negative deficit
angle along the $\xi_4$ direction, which can be interpreted as the black holes
being pushed apart by a ``rod" of mass per unit length
$\mu=1-\lambda_4/\lambda_3$ (which is negative).
For a general choice of $\alpha$, there will be
nodal singularities on both sides.  The mass per unit length of the outer
singularity will always be greater than that on the inside.

There is a degenerate case when the metric is free from nodal singularities.
Letting $r_+A=2/(3\sqrt 3)$ the roots $\xi_2$ and $\xi_3$ of the cubic
$\bar G$ become coincident. In this limit the range of $x$ becomes $\xi_3<
 x\le\xi_4$
since the proper distance between $\xi_3$ and $\xi_4$ diverges. The point
$x= \xi_3$ disappears from the $(x,\varphi)$ section which is no longer
compact but becomes
topologically $\R^2$, the sphere gaining an infinitely long tail.
One can eliminate the nodal singularity
at $x=\xi_4$ by choosing $\alpha=4\pi/\lambda_4$. It might seem that the
acceleration and black hole horizons become coincident in this limit. This is
not the case, however. The proper distance between the horizons (at fixed $x$
 and $t$) tends to a constant as $r_+A \rightarrow 2/(3\sqrt 3)$.
This case will be discussed further in section 4.

\newsec{Generating Dilaton Ernst}

\subsec{Generating new solutions}

In the case of vanishing dilaton coupling ($a=0$), Ernst \ernst\ has shown
that the nodal singularities can be removed by including a magnetic field
of the proper strength running along the symmetry axis.  The magnetic field
provides the force necessary to accelerate the black holes.
The magnetic field can be added to the C-metric via an Ehlers-Harrison
type transformation \he,
which takes an axisymmetric solution of the Einstein-Maxwell
equation into another such solution.  The same transformation applied to flat
spacetime produces Melvin's magnetic universe \melvin , which is the closest
one can get to a constant magnetic field in general relativity.
To follow the same path as Ernst, we first need to generalise the solution
generating technique that he employed to dilaton gravity.

In the case of Kaluza-Klein theory ($a=\sqrt{3}$), this turns out to be quite
simple.  It is known that the charged black holes in Kaluza-Klein theory can
be generated from
the uncharged ones ({\it i.e.} Schwarzschild with an extra compact spatial
dimension) by applying a coordinate transformation
mixing the time and internal
coordinates, a ``boost", and then
re-identifying the new internal coordinate \kkbh.
Similarly, we can add magnetic field
along the symmetry axis of an axisymmetric
solution to Kaluza-Klein theory by doing a transformation mixing the
internal and azimuthal coordinates, a ``rotation".
Applied to flat space, this transformation
reproduces the dilaton Melvin solution given in \gm\ for $a=\sqrt{3}$.
Together with the known form of the transformation without the dilaton field,
the Kaluza-Klein case provides sufficient
clues to guess the correct transformation
for general dilaton coupling.  In the string theory
case ($a=1$) this turns out to be one of
the $O(1,2)$ transformations\foot{
To obtain the full
$O(1,2)$ group one needs to include the antisymmetric tensor field
in the action \action.} that is known to act on
the space of axisymmetric solutions
\senny.

Let $(g_{\mu \nu}, A_\mu, \phi)$ be an axisymmetric
solution of \eom .  That is, all the fields are independent
of the azimuthal coordinate $\varphi$. Let the other three
coordinates be denoted by $\{x^i\}$. Suppose also
that
$A_i=g_{i\varphi}=0$.\foot
{We can relax this
condition and construct new solutions
only assuming axisymmetry. However, the
transformations are somewhat more involved
and will not be needed for our purposes.}
Then a new solution of \eom\ is given by
\eqn\gen{
\eqalign{
&g^\prime_{ij}=\Lambda^{2\over 1+a^2}g_{ij},\qquad
g^\prime_{\varphi\varphi}=\Lambda^{-{2\over 1+a^2}}g_{\varphi\varphi},\cr
&e^{-2a\phi^\prime}=e^{-2a\phi}\Lambda^{2a^2\over 1+a^2},\qquad
A^\prime_\varphi=-{2\over (1+a^2)B\Lambda}(1+{(1+a^2)\over
2}B A_\varphi),\cr
&\qquad
\Lambda=(1+{(1+a^2)\over 2}B A_\varphi)^2+{(1+a^2)\over 4}B^2
g_{\varphi\varphi}
e^{2a\phi}\cr}
}
The proof is presented in appendix A.
This transformation generalises the ``rotation"-type transformation
of Kaluza-Klein theory to all $a$. Although we will not write the explicit
transformations, it is possible
to generalise the ``boost"-type of transformation to all $a$, also.

\subsec{Dilaton Melvin}
Applying these transformations to flat space in cylindrical
coordinates we obtain the dilaton Melvin solutions given by Gibbons and
Maeda \gm ,
\eqn\dmelv{
\eqalign{
&ds^2=\Lambda^{2\over 1+a^2}\left[-dt^2+d\rho^2+dz^2\right]
+\Lambda^{-{2\over 1+a^2}}\rho^2d\varphi^2\cr
&e^{-2a\phi}=\Lambda^{2a^2\over 1+a^2},\qquad
A_\varphi=-{2\over (1+a^2)B\Lambda}\cr
&\qquad \Lambda=1+{(1+a^2)\over 4}B^2\rho^2\cr}
}
The parameter $B$ gives the strength of the magnetic field on the axis
via $B^2 = \left.{1\over 2} F_{\mu\nu} F^{\mu\nu}\right|_{\rho = 0}$.

\subsec{Dilaton Ernst}
Applying the transformation \gen\ to the dilaton C-metric finally yields the
dilaton Ernst solution.
\eqn\dernst{
\eqalign{
&ds^2=(x-y)^{-2}A^{-2}\Lambda^{2\over 1+a^2}
\left[F(x)\left\{G(y)dt^2-G^{-1}(y)dy^2\right\}
+F(y)G^{-1}(x)dx^2\right]\cr &\qquad +
(x-y)^{-2}A^{-2}\Lambda^{-{2\over 1+a^2}}F(y)G(x) d\varphi^2\cr
&e^{-2a\phi}=\Lambda^{2a^2\over 1+a^2}{F(y)\over F(x)},\qquad
A_\varphi=-{2\over (1+a^2)B\Lambda}(1+{(1+a^2)\over 2}Bqx)\cr
&\Lambda=(1+{(1+a^2)\over 2}Bqx)^2+{(1+a^2)B^2\over 4A^2(x-y)^2}
(1-x^2-r_+Ax^3)(1+r_-Ax)\cr}
}
Defining $\CG(y,x)=\Lambda^{-{2\over 1+a^2}}G(x)$ the nodal singularities
of the C-metric will be removed if
the period of $\phi$ is chosen to be $4\pi/|\partial_x\CG|_{\xi_3}$
and we impose $|\partial_x\CG|_{\xi_3} =|\partial_x \CG|_{\xi_4}$.
In the limit $r_\pm A\ll 1$, this constraint
yields Newton's law
\eqn\con{
mA\approx Bq,
}
where $m$ and $q$ are identified in terms of $r_+$ and $r_-$ according to
\mass .

Note that we can read off the dilaton Melvin metric in ``accelerated"
coordinates from \dernst\ by setting $r_+=r_-=0$.
The metric functions then reduce to
\eqn\melvinagain{
F=1,\qquad G(x)=1-x^2,\qquad
\Lambda=1+{(1+a^2)B^2\over 4A^2(x-y)^2}(1-x^2) .}
This form of the dilaton Melvin solution is useful for studying in what sense
the dilaton Ernst solution approaches dilaton Melvin.  This is discussed in
detail in Appendix B.  Here we note that,
if the value of the physical magnetic field parameter $B_p$ is defined
to be $\sqrt{{1\over 2} F_{\mu\nu}F^{\mu\nu}}$ on the axis as
$y\rightarrow \xi_3$,
\eqn\matching{B_p=\half
\left.{\partial_xG\over\Lambda^{3\over2}}\right |_{x=\xi_3} B_e.
}
where $B_e$ is the parameter that appears in \dernst. This value
of $B_p$ is also the amount of flux per unit area across a small area
transverse to the axis in the limit $y\rightarrow \xi_3$ as shown in
Appendix B.
In the limit $r_\pm A\ll 1$  this reduces to $B_p=B_e$.
Further, as discussed in Appendix B, we find coordinates in which the dilaton
Ernst metric \dernst\
approaches the dilaton Melvin metric
\dmelv\ near the outer axis for $r={1\over x-y}\rightarrow\infty$.

\newsec{Dilaton Instantons}

The above solutions describe two dilaton black holes accelerating
away from each other along the axis of a dilaton Melvin magnetic universe.
Euclideanising \dernst\ by setting $\tau=it$, we find that, just
as in the $a=0$ case, another condition must be imposed
on the parameters
in order to obtain a regular solution. The condition arises in order to
eliminate conical singularities at both the black hole and
acceleration horizons with a single choice of the period of $\tau$.
This is equivalent to demanding that the
Hawking temperatures of the two horizons are equal.

In terms of the metric function $G(y)$ appearing in \dernst, the period
of $\tau$ is taken to be $4\pi/|G^\prime(\xi_2)|$ and
the constraint is
\eqn\regular{
\left| G^\prime(\xi_2)\right| =\left| G^\prime(\xi_3)\right| , }
yielding
\eqn\roots{
\left({\xi_2-\xi_1\over \xi_3-\xi_1}\right)^{1-a^2\over 1+a^2}(\xi_4-\xi_2)
(\xi_3 -\xi_2)
=(\xi_4 - \xi_3)(\xi_3 -\xi_2). }
Recall that we have restricted our parameters so that
$\xi_4>\xi_3\ge\xi_2>\xi_1$.  For all values of $a$, \roots\
can be solved by demanding that the horizons have zero temperature
i.e. $\xi_2=\xi_3$.
We will refer to these
as type I instantons.
For $a<1$, the first factor on the left hand side
of \roots\ is
smaller than one, and there is also a second solution, which we will call
type II instantons.  For $a\ge 1$, the
first factor on the left is greater than one, corresponding
to the temperature of the black hole horizon always being
greater than the temperature of the acceleration horizon,
and there are no other solutions.

We first consider the type II solutions with $a<1$.  These generalise the
regular euclidean metrics considered in the $a=0$ case
\refs{\gwg, \garstrom, \andy}; the condition \roots\ is the analogue of the
$q=m$ condition on the parameters discussed in those papers.
In the limit $r_\pm A<<1$, for $a<1$ one can show that
the condition \roots\ leads to $r_+=r_-$.
Since we have chosen the parameters so that there are no nodal singularities
on the $t,y$ and $x,\varphi$ spheres, it is clear that the topology
of these spacetimes is $S^2\times S^2-{\rm\{pt\}}$ where the removed point
is $x=y=\xi_3$.

This instanton is readily interpreted as a bounce: the surface defined by
$\tau=0$ and $\tau=\pi$ has topology
$S^2 \times S^1 - {\rm{pt}}$, which is that
of a wormhole attached to a spatial slice of Melvin and is the zero momentum
initial data for the lorentzian ernst solution. In addition
the solution tends to
the Melvin solution at euclidean infinity (see appendix B).
The bounce describes the pair creation of a pair of oppositely charged dilaton
black holes in a magnetic field which subsequently uniformly accelerate
away from each other. From the metric
we deduce that there is a horizon sitting inside
the wormhole throat, located at a finite proper distance from the mouth.

Turning to the type I instantons, we note that this is again
the case of two coincident
roots, which has been discussed above in section 2.3 for $B=0$. There it was
pointed out that there are no nodal singularities even
in the absence of a magnetic field, and that consequently there
is no restriction on the value of $B$. The apparent coincidence of the
two horizons is an artifact of a poor choice of coordinates
and it can be shown
that the proper distance between the horizons remains finite as the
roots become coincident. Below, we exhibit a coordinate change,
originally used by Ginsparg and Perry to study Schwarzschild-DeSitter
instantons \ginsperry,  which
makes this explicit for the C-metrics.

In \C\ let $r_+ A = 2/(3{\sqrt 3}) - \epsilon^2/{\sqrt 3}$,
so that the limit of coincident
roots is $\epsilon \rightarrow 0$. Specifically, introducing
$y_0 = - \sqrt 3 (1 + {7\over 6} \epsilon^2)$ the roots to order $\epsilon^2$
are
\eqn\three{\eqalign{
&\xi_{2,3} = y_0\mp {\sqrt 3}\epsilon\cr
&\xi_4={{\sqrt 3}\over 2}(1+{1\over 6} \epsilon^2).\cr
}}
Writing the dilaton C-metric \C\ in the coordinates
\eqn\change{\chi = \cos^{-1}\left({1\over {\sqrt 3}\epsilon}\left(
y - y_0 \right)\right), \quad \psi = \sqrt 3 \epsilon t}
and taking the
limit $\epsilon\rightarrow 0$, then gives
\eqn\limitC{
\eqalign{
&ds^2=A^{-2}(x+ \sqrt3)^{-2}
\Biggl[F(x)\left\{- F(-\sqrt 3)^{{1-a^2}\over{2a^2}}\sin^2\chi d\psi^2
+ F(-\sqrt 3)^{{a^2-1}\over{2a^2}}d\chi^2
\right\}\cr
&\qquad\qquad+
F(-\sqrt 3)\left\{G^{-1}(x)dx^2+G(x)d\varphi^2\right\}\Biggr],\qquad
\qquad A_\varphi=qx,\cr
&e^{-2a\phi}={F(-\sqrt 3)\over F(x)},\qquad
G(x)= -{2\over{3\sqrt 3}}(x+\sqrt 3)^2(x-\sqrt 3/2)
(1+r_-Ax)^{1-a^2\over1+a^2}.\cr
}}

We may then apply the solution generating transformations
\gen\ with arbitrary parameter $B$, though we will not do this explicitly here.
Euclideanising \limitC\ by setting $\Psi = i \psi$, we see that
it is possible to
eliminate the conical singularities at the north and south poles
of the $(\Psi, \chi)$ section by making $\Psi$ periodic with period
$2\pi \left(1 - \sqrt 3 r_- A \right)^{{{(a^2 -1)}\over{(2a^2)}}}$.
Indeed, the $(\Psi, \chi)$ section is a round sphere and the topology of
this solution is $S^2 \times \R^2$, in contrast to the type II instantons.
It is not clear what, if any, the physical significance of these instantons
may be.

\newsec{Discussion}
The instantons presented in section 4
suggest that
topology changing processes can occur in dilaton gravity for $a<1$.
Specifically, the type II instantons describe the pair
creation in a uniform magnetic field
of an oppositely charged pair of $a<1$ dilaton black holes.
The rate of production of these black holes can be estimated
in the semi-classical approximation by calculating the action
\dggh.

It is interesting that the type II instantons exist only for
$a<1$. The interpretation of the type I instantons
which exist for all $a$ is unclear, especially since they exist for
any value of the magnetic field. It therefore appears
to be difficult to estimate the production rate of charged
black holes in theories with $a\ge 1$ using semi-classical techniques.

Including additional matter fields may yield
one way of modifying the type II solutions to obtain instantons
for $a\ge1$. In particular, in \dgt\ it was argued that in the
Einstein-Maxwell-Higgs theory which admits cosmic strings,
euclidean solutions exist which correspond to a string
world sheet wrapped around the horizon of a black hole.
The effect of the string is to cut out a ``wedge'' from the
$(r,t)$ section of euclidean Schwarzschild, an effect which could be
approximated in a vacuum theory by allowing a conical singularity
at the horizon with a specified deficit. Similarly, for $a=1$ say,
cosmic strings could be added to the model \action, in which case the type
II instanton with a certain conical singularity in the $(t,y)$ section
could describe the (cosmic string induced) pair creation
of $a=1$ black holes.

Another interesting possibility is that the physics of black hole pair
production for $a\ge 1$ is not so simply related to regular euclidean
instantons.
In reference \wilc\ it was shown that the thermodynamic behavior of charged
black holes with $a> 1$ differs from that given by
the naive interpretation of their euclidean sections.
For $0\le a<1$ the temperature of the extremal black holes goes to zero and
Hawking radiation is extinguished, as one would expect.
For $a>1$, however, the
temperature of a black hole,
given by the periodicity of its euclidean section,
diverges as extremality is approached, a result that was regarded as puzzling
\ghs .
This puzzle was partially resolved in \wilc , where it was shown that,
for $a>1$, the Hawking radiation is in fact shut off by infinite grey
body factors.
For $a=1$ the temperature of the extremal black hole
approaches a constant and the results of
the analysis in
\wilc\ are inconclusive.  In our case, demanding
regularity of the euclidean section of
the dilaton Ernst metric is equivalent to requiring that the black hole be in
thermal equilibrium with the acceleration radiation.
It is tempting to suspect that the
inability to achieve this (for nonzero temperature) for $a\ge 1$
is somehow related to the physics uncovered in \wilc .
The study of black hole pair production with $a\ge 1$ would then
require more subtle methods.

In the introduction, we summarised the cornucopion scenario for
resolving
the paradoxes associated  with information loss in the
scattering of particles from extremal $a=1$ dilaton black holes.
In this scenario it is important
that the extremal black hole is non-singular, with
an infinitely long
throat leading to a second null infinity. Since the
extremal black hole geometries (using the total metric \total) for
$0<a<1$ also have this property, it is natural to conjecture that
these models all admit cornucopion type scenarios.
Moreover, the instantons we
have constructed above indicate that for $0<a<1$ there may not be
a problematic
infinite pair production
of cornucopions in a magnetic
field.
Firstly, we expect the action of the instanton to be finite.
Furthermore,
since the created wormholes have finite length it would appear that
if one included matter fields and calculated the one-loop
determinants, one would not be including the infinite number of
states living far down the static wormholes\foot{We note however, that
this logic has been questioned in ref. \sg.}. In conclusion,
one expects the pair production rate of cornucopions to be finite since
a cornucopion is not an elementary particle but is deeply
interconnected with the geometry of spacetime.
As we have noted, we cannot say anything definite about the case $a=1$.
It would be interesting to understand the implications of our exact results
for the approximate instantons
presented
in \banksol.

\bigskip
{\bf Note Added}: We note that the dilaton Ernst solution \dernst\
does not asymptote to the dilaton Melvin solution \dmelv\ although
the metrics do match. A comparison of the gauge field and dilaton
for \dernst\ and \dmelv\ shows that the solutions are related by
a constant shift in the dilaton and rescaling of the gauge field
as in footnote 1. The details of this are given in \dggh.

\bigskip\centerline{\bf Acknowledgements}\nobreak
We would like to thank David Garfinkle, Steve Harris,
Jeff Harvey, Gary Horowitz, Gary
Gibbons, Martin O'Loughlin and Bob Wald for useful discussions.
We especially thank Robert Caldwell for his help with
Mathematica \MATHEMATICA\ and MathTensor \MATHTENSOR\ which we used
to check the C-metric solutions.
DK and JT
thank the Aspen Center
for Physics for its hospitality during part of this work.
JPG is supported by a grant from the
Mathematical Discipline Center of the Department of Mathematics,
University of Chicago. JT is supported in part by NSF grant
NSF-THY-8714-684-A01 and FD by the DOE and
NASA grant NAGW-2381 at Fermilab.

\appendix A ~
Suppose that we have a solution to \eom\ that
is axisymmetric, i.e. independent of the azimuthal coordinate
$\varphi$,  and further satisfies $A_i=g_{i\varphi}=0$, where
$x^{i}$ are the other three coordinates.
We prove that the transformations \gen\ generate a new solution by
showing that the the transformations leave the action \action\ invariant.
We first rewrite the action in terms of the rescaled total metric
\total\
to obtain
\eqn\act{S=\int d^4x{\sqrt {-g_T}}e^{-2\phi/a}\left[
R_T+\left({6-2a^2\over a^2}\right)(\nabla\phi)^2-e^{2\phi{{1-a^2}\over a}}
F^2\right]
}
where indices are raised with the inverse of the total metric.

Introducing the definitions
\eqn\defs{
\eqalign{{}^3\! g_{ij}&=g_{Tij},\qquad\qquad V=g_{T\varphi\varphi}\cr
\tilde\phi&=\phi-{a\over 4}{\rm log}V,\cr}
}
we can recast the action into the form
\eqn\actt{
\eqalign{
S=\alpha\int d^3x{\sqrt{-\,{}^3\!g}}&e^{-{2\tilde\phi/a}}\big[
{}^3\!R+{6-2a^2\over a^2}\partial_i\tilde\phi\partial^i\tilde\phi
-{a^2-1\over a}V^{-1}\partial_i\tilde\phi\partial^i V\cr
&-{1+a^2\over 8}V^{-2}\partial_i V\partial^i V
-2e^{2\tilde\phi{{1-a^2}\over a}}V^{-{1+a^2\over 2}}\partial_i
 A_\varphi\partial^i
A_\varphi\big]
\cr}
}
where we have carried out the integration over $\varphi$ assuming
its range is $0\le\varphi\le \alpha$.
The virtue of these definitions is that the transformations \gen\
now take the simple form
\eqn\gent{
\eqalign{
{}^3\!g^{\prime}_{ij}&={}^3\!g_{ij}\cr
V^\prime&=\Lambda^{-4\over 1+ a^2}V\cr
\tilde\phi^\prime&=\tilde\phi\cr
A_\varphi^\prime&=-{2\over (1+a^2)B\Lambda}(1+{(1+a^2)\over 2}BA_\varphi)\cr}
}
To complete the proof a straightforward calculation shows
that the Lagrangian
is invariant under these transformations.

\appendix B ~

\subsec{Determination of the magnetic flux}

Here we will determine the value of the physical magnetic
field parameter, $B_p$.
The magnetic flux through a small area around the
axis is given by
\eqn\flux{
\eqalign{
\Delta {\rm flux} = \int da_{\hat{y}}B^{\hat{y}}& = \int d\varphi
dx\partial_x A_\varphi
\cr
&\simeq\left . {\partial A_\varphi\over\partial x}
\right |_{x=\xi_3}\Delta \varphi\Delta x \cr} }
The flux per unit area in the limit $y\rightarrow \xi_3$ in dilaton Ernst
is then given by
\eqn\fluxarea{ B_p =
\left.{\Delta {\rm flux}\over\Delta {\rm area}}\right| _{x=\xi_3} =
\half\left.{\partial_x G\over\Lambda^{{3\over 2}}}\right|_{x=\xi_3}
B_e.
}
and in the limit $r_\pm A\ll 1$ reduces to $B_p=B_e$.
It is the same relation one gets from
just matching $F_{\mu\nu}F^{\mu\nu}$ on the axis.

\subsec{Matching the metric near the axis}

We now show that the dilaton Ernst metric
approaches the dilaton Melvin metric
near the outer axis as
$r\rightarrow\infty$.
Note that for the euclidean section the outer axis,
$x,y\rightarrow\xi_3$, is the only place where $r\rightarrow\infty$.
We start with the dilaton Melvin metric expressed in
accelerated coordinates, as discussed in section 3.3.  There is then an
acceleration parameter $\bar{A}$ at our disposal in the matching.
Near the axis
$G(x)\simeq\lambda_3(x-\xi_3)$, where $\l3 = \partial_x G|_{x=\xi_3}$.
Make the following coordinate transformation
in the dilaton Ernst
metric \dernst\
\eqn\rescale{
\bt =\half\lambda_3 t,\qquad \by =-{y\over\xi_3}, \qquad
\bphi = {\l3 \varphi\over 2\Lambda(\xi_3)^{2\over 1+a^2}},\qquad \brho
=\left( {2(\xi_3-x)\over\xi_3}\right )^\half .
}
Near the axis the dilaton Ernst metric then has the form
\eqn\ernstaxis{
ds^2_{\rm ernst}\simeq
{-2\xi_3 F(\xi_3)\Lambda(\xi_3)^{2\over 1+a^2}\over
\l3 (1+\by^2) A^2\xi_3^2}
\left[
2(\by +1)d\bt^2 - {d\by^2\over2(1+\by)} + \brho^2d\bphi^2 + d\brho^2\right] }
We can make a similar coordinate transformation on the accelerated form of the
dilaton Melvin metric near the axis to get

\eqn\melvinaxis{
ds^2_{\rm melvin}\simeq{1\over \bA^2(1+\by)^2}\left [
2(\by +1)d\bt^2 - {d\by^2\over2(1+\by)} + \brho^2d\bphi^2 + d\brho^2\right] }
These two are the same if we identify the acceleration of the Melvin coordinate
system $\bar{A}$ according to
\eqn\accel{
{1\over\bA^2}=-{2\over A^2\l3\xi_3}F(\xi_3)\Lambda(\xi_3)^{2\over 1+a^2} }
Since the choice of $\bar{A}$ is just a choice of coordinates,
this shows that the two metrics are the same near the axis.

We
note that in the preceding calculation,
we took the limit
$x,y\rightarrow\xi_3$ in the manner
\eqn\limit{
x-\xi_3\rightarrow 0,\qquad y-\xi_3\sim (x-\xi_3)^q,\qquad q<\half }
which lets $r^2 G(x)\rightarrow 0$ on the axis.  Choosing $q>\half$ gives an
artificially singular slicing of the spacetime.
Taking the
limit for $q=\half$ shifts the constant $\Lambda(\xi_3)$.



%
%
%
%
%

\listrefs
\end